# Strategy to Nonlinearly Deplete Irreversible Li Consumption in Si-rich Li-Ion Batteries

K.Ogata,[1,2,z] K.Takei,[1] S.Saito,[2] S.Wakita,[1] M.Koh,[1] SG. Doo,[1] S.Han,[1] S.Jeon,[1 z]

Affiliation(s):

1. Samsung Advanced Institute of Technology (SAIT), Samsung Electronics, Samsung-ro 130, Suwon, Gyeonggi-do, 16678, Korea

2. Samsung Research Institute of Japan (SRJ), Samsung Electronics, 2-1-11, Senba-nishi, Mino-shi, Osaka-fu, 562-0036, Japan

[z]Corresponding Author E-mail Address [k.ogata@samsung.com], [seongho.jeon@samsung.com]

**Abstract**

Despite recent significant developments of Si composites, use of silicon with significance in the anodes for Li-ion batteries is still limited. In fact, nominal energy density is to be saturated around ~750 Wh/L regardless of cell-types under the current material strategies. Use of Si-rich anode can push the limit; however, the prolonged irreversible Li consumption becomes more prominent. We previously showed that repeating c-Li$_{3.75(+\delta)}$Si formation/decomposition, typically recognized to degrade the anodes, can improve the irreversibility and accumulatively minimize the gross consumption. Utilizing the insights combined with prelithiation techniques, here we provide prototypic cell designs that can nonlinearly deplete the consumption.

**Main text**

**Introduction**

Si has been attractive alternative to Gr as the negative electrode in the Li-ion battery (LIB), owing to its significantly higher specific capacity (~3579 mAh g$^{-1}$, assuming Li$_{3.75}$Si at ambient temperature). Nominal density in cutting-edge LiB cells for larger applications typically lies around ~700 Wh/L with NCA/NCM (811/622) coupled with Gr (optionally blended with partly prelithiated 3–5w % $SiO_x$). However, a learning curve of the density with the current material strategies, i.e. a fraction of Si in Gr with higher current density (mAh/cm$^2$) e.g. >5 mA/cm$^2$, gets sluggish over the last 5–10 years and is most likely to be capped around ~750 Wh/L regardless of cell types. In contrast, Si-*rich* anodes can present one of very few remaining options to certainly leap the density, being able to reach above 900 Wh/L and can also undergo fast charging owing to prominent Li diffusivity in Si bulk. Despite the excellency, such anodes have not emerged on the market. One of the most critical bottlenecks in pragmatic cells is output capacity loss via prolonged irreversible Li consumption in the Li–Si processes, which is often quantified by Coulombic efficiency (CE, the delithiation/lithiation capacity ratio);[1-4] note that sustaining reversible anode capacity retention is rather handled by elaborated various Si-C composites [5] but CE is not particularly when Si concentration in the anode and its current density (mAh cm$^{-2}$) gets higher, e.g. > 500 mAh g$^{-1}$ and > 5 mAh cm$^{-2}$. We previously showed that repeating c-$Li_{3.75(+\delta)}$Si formation and decomposition,[6] which is typically recognized as one of degradation factors in the anodes,[7-10] has a capability to improve CE and consequently minimize the cumulative irreversible Li consumption. Utilizing the insights, here we provide prototypic cell designs for the first time that can *in situ* nonlinearly deplete the Li irreversible consumption over cycling.

**Experimental**

The anode is designed so that Li–Si process can undergo abrupt phase transformations (a-$Li_{3.5–3.75}$Si →c-$Li_{3.75(+\delta)}$Si) in proportion to the stoichiometry even at higher current rates, e.g. > 1C. Active materials in the form of secondary particles are synthesized by conventional spray-drying method (B290 Mini Spray-dryer, Buchi). These secondary particles are composed of defective poly-crystalline (pc-)Si nano-powder (pc-SiNP, Stream-Si, ~120 nm), multi-wall carbon nanotubes (MWCNT, 15 nm, CNT Co. Ltd.) with flake-type Gr (FT-Gr, SPG1, SEC carbon), and polyvinyl alcohol (PVA, Sigma Aldrich, MW ~50 k).[6] Defective pc-SiNP is used because the defective sites (twin boundaries and stacking faults) work as catalysis to accelerate the phase transformation.[6] First, the components are dispersed in DI water, followed by 2 h of ultrasonication. The SiNP/MWCNT/FT-Gr/PVA ratios are 55.9/7.8/34.3/2.0 (wt%), theoretically and experimentally being ~2273 and 2250 mAh g$^{-1}$, respectively. The secondary particles are porous, which enable prompt wetting to electrolyte.[11] The secondary particles and Gr (SFG6, Timrex and MC20, Mitsubishi) are blended to be an anode capacity of 760 mAh g$^{-1}$. The electrode consists of 20 wt% secondary particles and 71 wt% Gr as active materials, 8 wt% polyacrylic acid (Li-PAA, Hwagyong Chemical) as a binder, and 1 wt% Kechen Black (KB) as a conductive additive. The cathode consists of 94 w% NCM622 (Tanaka), 5 w% of PVdF as a binder (Sigma Aldrich), and 1w% KB, being and 2.44 mAh cm$^{-2}$. 2032-Type coin full cells (Hohsen Corp.) are used in all the following experiments. The electrolyte is 1 M LiPF$_6$ in a 25/5/70 (vol%) mixture of fluoroethylene carbonate (FEC)/dimethyl carbonate (EC)/dimethyl carbonate (DEC) (LP 30 Selectilyte, Merck). A 10-µm-thick separator (Celguard, Asahi Kasei) is used.

Ogata *et.al* previously uncovered [6] that repeating the phase transformation, often recognized as a degradation factor in the anode, can improve CE and minimize cumulative irreversible Li consumption. However, as widely recognized, the reactions involves sacrificial capacity degradation and irreversible Li consumption.[6] Accordingly, the prototypic cell (Fig. 1a-c) presented here is designed to carry extra capacity loading and prelithiated Li-dose in the anode, which are to be sacrificially consumed over cycling in exchange of implementing significantly high reversibility in the anode.[6] We define conventional cell design to have the negative positive capacity ratio (NP ratio) of 1.05 (2.56/2.44 mAh/cm$^2$ for anode/cathode, respectively) as shown in Fig. 1a. In contrast, the prototypic cell is designed to have NP ratios of 1.25–1.58 (i.e. anode current density ranging from 3.05 to 3.86 mAh cm$^{-2}$) with a prelithiated Li-dose of 15–50 % of the anode capacity (0.6–1.95 mAh cm$^{-2}$) as shown in a lower part of Fig. 1a. Thus, under the prototypic design, state of charge (SOC) after charging in the anode can reach ~100% or slightly higher since the initial vacant capacity in the anodes is smaller than cathode capacity. This enables the anode to undergo the phase transformation on charging (a-Li$_{3.5-3.75}$Si →c-Li$_{3.75+\delta}$Si) and accordingly deplete the irreversible consumption as shown in Fig. 1b. For the prototypic design, the anode was preliminary prelithiated in a separate Li-metal countered coin cell. Subsequently, the prelithiated cells are disassembled in Ar-purged glovebox, and reassembled into the coin full cells. In the following experiments, the full cells are typically cycled at 0.1/0.2C (1$^{st}$/2$^{nd}$ cycle) and 1 C (>3$^{rd}$ cycle) under constant current constant voltage (CCCV) on charging (CV cutoff at C/100) and constant current on discharging from 2.8 to 4.2 V.

One apparent note is that such combination of NP ratio and prelithiation is typically unacceptable in the production consensus for the conventional Gr-rich electrodes (with a fractional Si/SiO$_x$), the capacity typically being e.g. 350–500 mAh g$^{-1}$. This is because being SOC100% or higher can yield Li dendrite on charging. In contrast, Si-rich anodes can more efficiently accommodate extra Li owing to +δ component in c-Li$_{3.75+\delta}$Si,[11] resulting in mush less susceptible to dendrite formation. Another important note is that in order to implement higher volumetric energy density (Wh/L), the specific capacity (mAh g$^{-1}$) in the anode needs to be set high enough to offset the energy density loss by the sacrificial loading; typically, the capacity of >700 mAh g$^{-1}$ is required for the density of >750 Wh/L under our cell configuration; the tabulated Wh/L presented here takes into account volume expansion upon anode prelithiation. Accordingly, use of the conventional Gr(–Si/SiO$_x$) anodes with a capacity range of e.g. ~380–500 mAh g$^{-1}$ is excluded under the nonlinear depletion strategies.

**Results and Discussion**

**Fig. 2** shows electrochemical outputs over cycles for the prototypic cell design; anode/cathode current densities are 3.32 and 2.44 mAh cm$^{-2}$ (NP~1.38), respectively, and the former is prelithiated by 1.41 mAh cm$^{-2}$ (a prelithiation dose of ~40% out of the anode capacity). As shown in Fig. 2a-c, the capacity and CE over cycling show much higher retention rate and higher values, respectively, compared to the conventional ones. However, this does not solely explain whether the prominent retention owes to the inherently *in situ* improved irreversibility nature or to merely compensating the consumed Li fractionally using prelithiated Li reservoir cycle by cycle. To preliminarily interpret nature of the prelithiation effects, another retention profile is presumably added to the chart (dotted blue line in Fig. 2a), supposing that a degradation rate follows the conventional one starting with the total capacity loading in the anode at the end of charging (i.e., including the prelithiated Li). The result shows that a parity cycle number, at which Li reservoir in the cell design presumably runs out, comes near 120$^{th}$ cycle. It is to be noted that even far after the parity cycle, the capacity and the CE in the prototypic design are sustained at higher levels compared to the conventional ones, the former being 81.5 % after 300 cycles and ~99.5%, respectively (Fig. 2b,c).

These results suggest that the inherent nature of irreversibility in the anode under the prototypic cell design is significantly improved over the early cycling stage, which is distinctive from the conventionally recognized benefit of prelithiation; supplying a portion of prelithiated Li reservoir in the anode cycle by cycle to compensate the irreversible consumption. On top of this, we for the first time highlighted a new feature of prelithiation that an appropriate combination of a NP ratio and a prelithiated Li-dose in the anode can nonlinearly deplete the irreversible Li consumption, the underlying mechanism of which probably lies in previously reported our insights; the CE improvement via repeating c-

Li$_{3.75+\delta}$Si formation and decomposition. [6]

As shown in Fig. 3, we parameterized NP ratio and doping dose in the anodes to capture the parameter space that leads to the nonlinear depletion of irreversible Li consumption. Conventional cell design is mapped on a solid orange line which sits on an elongated dashed orange line that represents non-prelithiated cell designs. Domain of prelithiated cell design lies underneath the dashed line, highlighted by green area, in which we enclosed two domains that does/does not reach 80% retention at 300 cycles by blue and red circles, respectively. The nonlinear depletion more efficiently occurs in certain parameter space of NPs and prelithiation doses. Further details will be presented elsewhere.

**Summary**

In this study, we provided prototypic cell design of Li-ion batteries that can *in situ* nonlinearly deplete the Li irreversible consumption in Si-rich anode. The cell was designed to undergo amorphous–crystalline phase transformation between a-Li$_{3.5–3.75}$Si and c-Li$_{3.75+\delta}$ to utilize recently reported our insights that the process can improve the reversibility and consequently minimize cumulative irreversible Li consumption. However, as often recognized, repeating the reaction involves certain sacrificial capacity loss and Li consumption. Accordingly, we for the first time provide a cell design that carries an extra capacity loading (NP~1.25–1.58) and a prelithiated dose in the anode (15–50 %), which are sacrificially consumed over cycling in exchange of significantly high reversibility. The strategy is distinguished from conventional prelithiation of Si that works as Li reservoir to compensate the irreversibility. Consequently, the prototype cell (~800 Wh/L) outperforms a conventional cell, the former being ~82% at 300$^{th}$ cycles.

# References

(1) Choi, N.-S.; Yew, K. H.; Lee, K. Y.; Sung, M.; Kim, H.; Kim, S.-S. *J. Power Sources* **2006**, *161*, 1254.
(2) Meyer, B. M.; Leifer, N.; Sakamoto, S.; Greenbaum, S. G.; Grey, C. P. *ECS Solid State Lett.* **2005**, *8*, A145.
(3) Nie, M.; Abraham, D. P.; Chen, Y.; Bose, A.; Lucht, B. L. *J. Phys. Chem. C* **2013**, *117*, 13403.
(4) Michan, A. L.; Divitini, G.; Pell, A. J.; Leskes, M.; Ducati, C.; Grey, C. P. *J. Am. Chem. Soc* **2016**, 7918.
(5) Liu, Y.; Zhou, G.; Liu, K.; Cui, Y. *Accounts of Chemical Research* **2017**, *50*, 2895.
(6) Ogata, K.; Jeon, S.; Ko, D.-S.; Jung, I.; Kim, J.; Ito, K.; Kubo, Y.; Takei, K.; Saito, S.; Cho, Y.; Park, H.; Jang, J.; Kim, H.; Kim, J.-H.; Kim, Y.; Koh, M.; Uosaki, K.; Doo, S.-G.; Hwang, Y.; Han, S.-s. *arXiv:[cond-mat.mtrl-sci]* **2017**, *1706.00169*.
(7) Hatchard, T. D.; Dahn, J. R. *J. Electrochem. Soc.* **2004**, *151*, A838.
(8) Kang, Y.-M.; Lee, S.-M.; Kim, S.-J.; Jeong, G.-J.; Sung, M.-S.; Choi, W.-U.; Kim, S.-S. *Electrochem. Commun.* **2007**, *9*, 959.
(9) Obrovac, M. N.; Christensen, L. *ECS Solid State Lett.* **2004**, *7*, A93.
(10) Obrovac, M. N.; Krause, L. J. *J. Electrochem. Soc.* **2007**, *154*, A103.
(11) Ogata, K.; Salager, E.; Kerr, C. J.; Fraser, A. E.; Ducati, C.; Morris, A. J.; Hofmann, S.; Grey, C. P. *Nat Commun* **2014**, *5*, 3217.

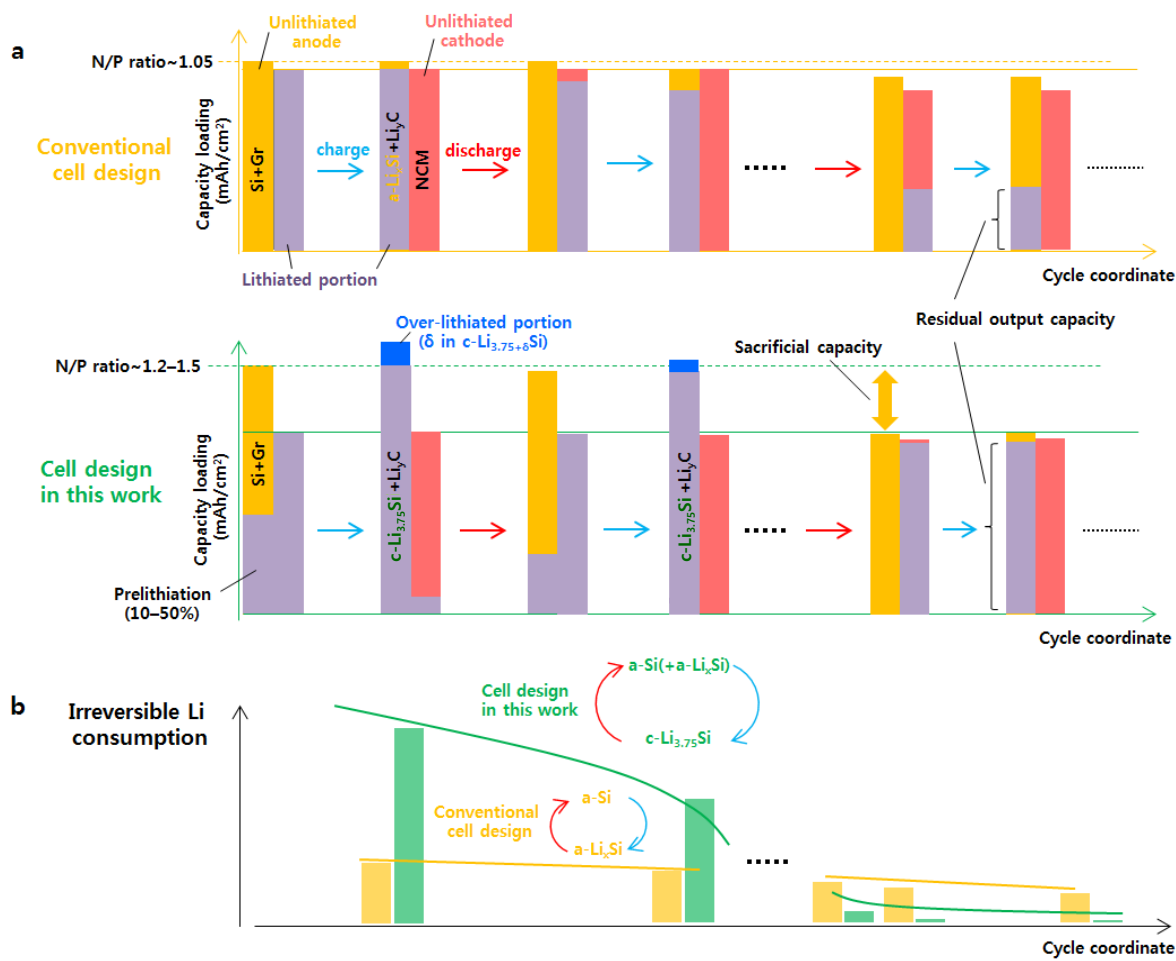

**Fig. 1 Strategy to in situ nonlinearly deplete irreversible Li consumption in Si-rich anodes**

(a) Schematics of conventional and prototypic cell designs that have a negative/positive capacity loading ratio of 1.05 and 1.2–1.5 (prelithiated by 15–50% of the anode capacity), respectively. Over cycling, Li–Si processes in the former and the latter designs are dominated by reactions between a-Si and a-Li$_x$Si and between a-Si/a-Li$_x$Si and c-Li$_{3.75}$Si.[6] (c) Schematics of irreversible Li consumption plotted over cycle number for the conventional and the prototypic cell designs.

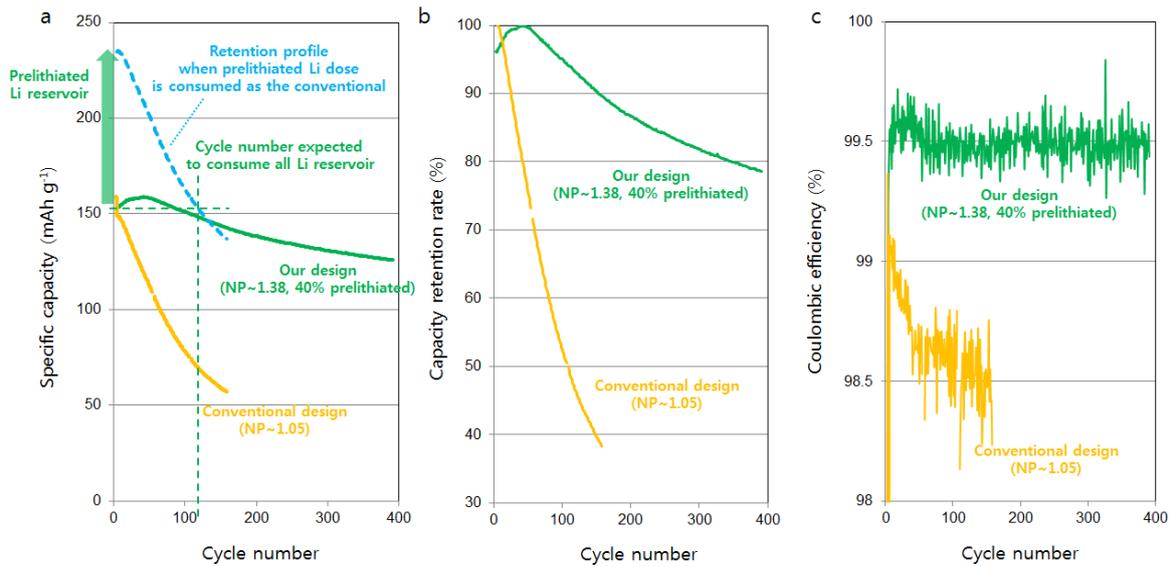

**Fig. 2 Electrochemical outputs for the two different cells**

(a) Specific capacity over cycle number for cells with our design (green) and conventional one (orange). A dashed blue line shows a presumed retention profile with our design given that the irreversible Li consumption rate follows that for the conventional cell. The specific capacity starts from the point that includes prelithiated Li dose in the anode. Parity cycle number, at which Li reservoir runs out under the assumption on the dashed blue line, is indicated by a cross section of dashed green lines. (b) Capacity retention rate and (c) Coulombic efficiency over cycle number for cells with our prototypic design (green) and conventional one (orange).

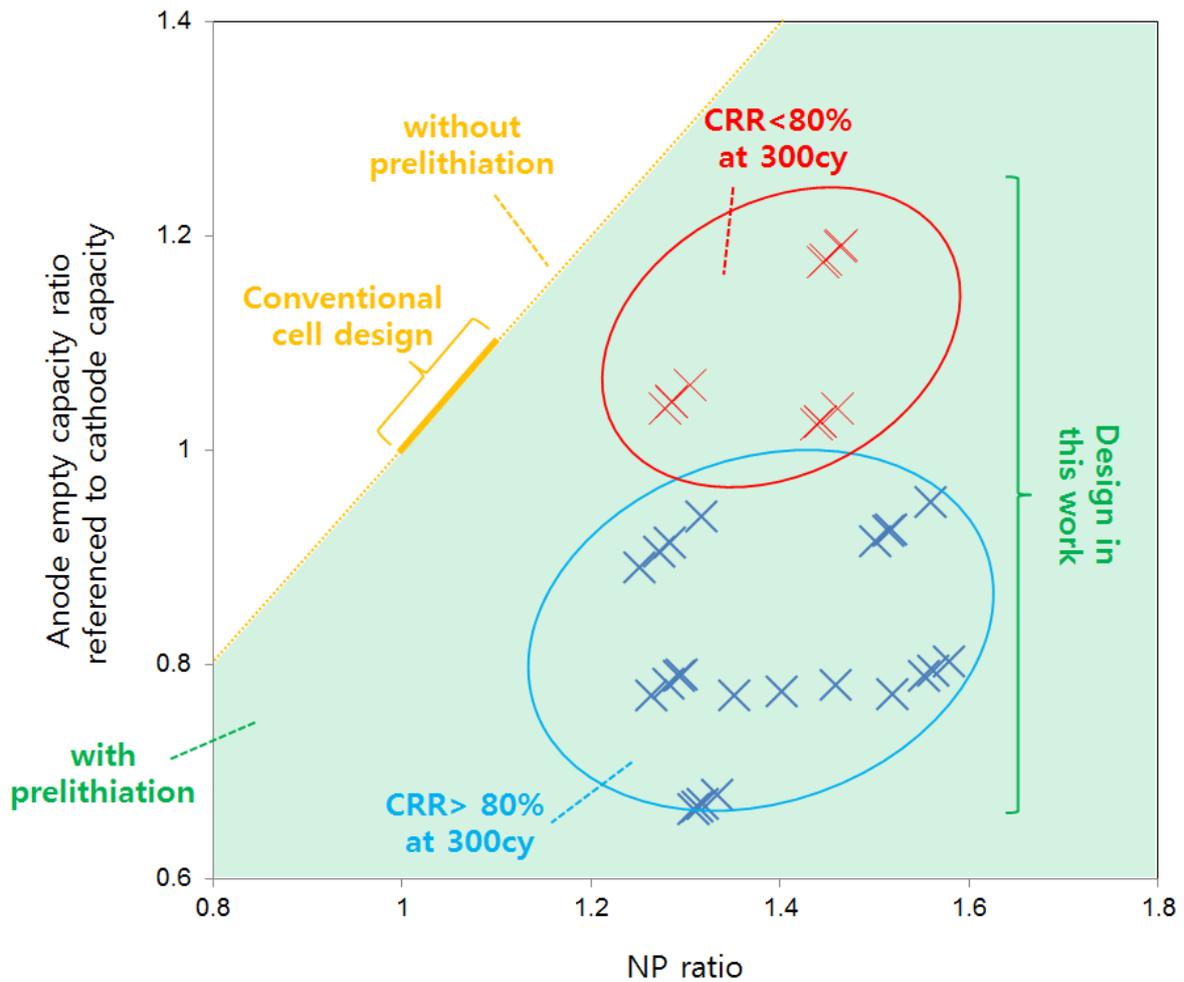

**Fig. 3 Negative/positive capacity loading ratio plotted over empty capacity ratio in anode referenced to cathode.**

Conventional cell design lies on a solid orange line, which sits on cell designs without prelithiation (a dashed orange line). Domain of prelithiated Si is mapped underneath the dashed orange line. The blue/red crossed plots enclosed by blue and red lines show clusters of the points that do and do not reach a capacity retention rate of 80% after 300 cycles, respectively.